\documentclass{elsart}
\usepackage{natbib}

\usepackage{graphics}
 \usepackage{graphicx}
 \usepackage{epsfig}
\usepackage{amssymb}

\begin{document}

\begin{frontmatter}


\title{Scaling Relation between Sunyaev-Zel'dovich Effect and X-ray
   Luminosity and Scale-Free Evolution of Cosmic Baryon Field}

\author[1,2]{Qiang Yuan}\ead{yuanq@mail.ihep.ac.cn}
\author[1]{Hao-Yi Wan}
\author[1,3,4]{Tong-Jie Zhang\corauthref{cor1}}\ead{tjzhang@bnu.edu.cn}
\corauth[cor1]{Corresponding author.}
\author[4]{Ji-Ren Liu}
\author[5,6]{Long-Long Feng}
\author[4]{Li-Zhi Fang}\ead{fanglz@physics.arizona.edu}

\address[1]{Department of Astronomy, Beijing Normal
University, Beijing, 100875, P.R.China}
\address[2]{Key Laboratory of Particle Astrophysics, Institute
of High Energy Physics, Chinese Academy of Sciences, 
Beijing 100049, P.R.China}
\address[3]{Kavli Institute for Theoretical Physics China,
Institute of Theoretical Physics, Chinese Academy of Sciences
(KITPC/ITP-CAS), P.O.Box 2735, Beijing 100080, P.R. China}
\address[4]{Department of Physics, University of Arizona,
Tucson, AZ 85721}
\address[5]{Purple Mountain Observatory, Nanjing 210008,
P.R. China}
\address[6]{National Astronomical Observatories, Chinese
Academy of Science, Chao-Yang District, Beijing, 100012, P.R. China}

\begin{abstract}

It has been revealed recently that, in the scale free range, i.e.
from the scale of the onset of nonlinear evolution to the scale of
dissipation, the velocity and mass density fields of cosmic baryon
fluid are extremely well described by the self-similar log-Poisson
hierarchy. As a consequence of this evolution, the relations among
various physical quantities of cosmic baryon fluid should be scale
invariant, if the physical quantities are measured in cells on
scales larger than the dissipation scale, regardless the baryon
fluid is in virialized dark halo, or in pre-virialized state. We
examine this property with the relation between the Compton
parameter of the thermal Sunyaev-Zel'dovich effect, $y(r)$, and
X-ray luminosity, $L_{\rm x}(r)$, where $r$ being the scale of
regions in which $y$ and $L_{\rm x}$ are measured. According to the
self-similar hierarchical scenario of nonlinear evolution, one
should expect that 1.) in the $y(r)$-$L_x(r)$ relation,
$y(r)=10^{A(r)}[L_{\rm x}(r)]^{\alpha(r)}$, the coefficients $A(r)$
and $\alpha(r)$ are scale-invariant; 2.) The relation
$y(r)=10^{A(r)}[L_{\rm x}(r)]^{\alpha(r)}$ given by cells containing
collapsed objects is also available for cells without collapsed
objects, only if $r$ is larger than the dissipation scale. These two
predictions are well established with a scale decomposition analysis
of observed data, and a comparison of observed $y(r)$-$L_x(r)$
relation with hydrodynamic simulation samples. The implication of
this result on the characteristic scales of non-gravitational
heating is also addressed.

\end{abstract}

\begin{keyword}
cosmology: theory \sep large-scale structure of universe \sep
X-rays: galaxies: clusters \sep hydrodynamics \sep methods: numerical

\PACS 95.30.Jx \sep 07.05.Tp \sep 98.80.-k
\end{keyword}

\end{frontmatter}

\section{Introduction}

Scaling relation of dimensional quantities is very powerful to
reveal the dynamical feature of various physical systems. There has
been a considerable effort devoting to study the correlations and
scaling laws of various observable quantities of galaxy clusters.
Since virialized self-gravitational system is characterized by one
parameter, mass or virial temperature, one can find a set of scaling
relations among mass, size, X-ray luminosity, temperature, and
Compton parameter of Sunyaev-Zel'dovich (SZ) effect if the velocity
and mass density fields of baryon fluid in clusters are assumed to
be similar to the virialized dark matter halos \citep{1986MNRAS.222..323K}.
Observed data of galaxy clusters did yield scaling relations
\citep{1991MNRAS.252..428E,1993ApJ...412..479D,1999ApJ...524...22W,
2000MNRAS.315..356H,2000ApJ...538...65X,2005MNRAS.357..279C}.
However, observed scaling relations generally do not support the 
predictions given by the baryon-dark matter similarity of virialized 
dark halos \citep{2000MNRAS.315..356H,2000MNRAS.315..689L}.

Since Newtonian gravity is scale-free, the self gravitational system
of collisionless dark matter shows scaling behavior if the power
spectrum of initial density perturbations is scale-free. These
scaling is regardless of whether the underlying gravitational field
is virialized \citep{1980lssu.book.....P}. Thus, if the velocity and
mass density fields of cosmic baryon matter are given by a similar
mapping of the fields of dark matter, one may expect the scaling
relations of clusters. However, the similar mapping assumption is
correct only in linear regime \citep{1992A&A...266....1B}, but is
baseless in nonlinear regime \citep{1989RvMP...61..185S}. The
nonlinear evolution of cosmic baryon fluid leads to statistically
decouple of the fluid from dark matter. The statistical properties
of the velocity and mass density fields of baryon fluid do show
deviation from the underlying dark matter field
\citep{2004ApJS..154..475P,
2005ApJ...623..601H,2005ApJ...625..599K}.

Nevertheless, it has been pointed out by \cite{1989RvMP...61..185S}:
the dynamics of cosmic baryon fluid in the expanding
universe is scale-free, i.e. no preferred special scales can be
identified in the range from the onset of nonlinear evolution down
to the length scale of dissipation. It likes fully developed
turbulence in inertial range. This idea recently received
substantial developments. With the hydrodynamic simulation sample of
the concordance $\Lambda$CDM model, the velocity field of cosmic
baryon fluid is found to be extremely well described by
She-Leveque's (SL) scaling formula \citep{1994PhRvL..72..336S} in the
``inertial range'' \citep{2006PhRvL..96e1302H}. The SL formula is
considered to be the basic statistical features of the scale-free evolution
of fully developed turbulence. Moreover, the SL formula comes from
self-similar log-Poisson hierarchy, which is related to the hidden
symmetry of the Navier-Stokes equations \citep{1994PhRvL..73..959D,
1995PhRvL..74..262S}. Very recently, it has been shown that the clustering
of the mass density field of the cosmic baryon fluid can indeed be
well described by a log-Poisson hierarchical cascade
\citep{2008ApJ...672...11L}. All the scaling relations and non-Gaussian
features predicted from the log-Poisson hierarchy are in very good
agreement with the hydrodynamic simulation samples.

These results indicate that, in the scale-free range, the nonlinear
evolution of cosmic baryon fluid reaches a statistically
quasi-steady state similar to a fully developed turbulence. For
turbulence of incompressible fluid, the fluid undergoes a
self-similar hierarchical evolution from largest to the smallest
eddies and finally dissipates into thermal motion. For cosmic baryon
fluid, the clustering on different scales can also be described by a
self-similar hierarchy, and the fluid finally falls and dissipates
into thermal motion.

This scenario motives us to investigate the scaling relations of
clusters from the self-similar hierarchy of cosmic baryon fluid. If
the observed scaling relations come from the self-similar hierarchy,
one can expect that 1.) the relations of dimensional quantities
should be scale-free, i.e. all the scale-dependent coefficients of
the scaling relations are scale-invariant; 2.) the relations should
be held only if the scales of considered regions are larger than
Jeans length, regardless of whether the underlying gravitational
field is virialized, i.e. the relations given by cells containing
collapsed objects is also available for cells without collapsed
objects, only if the scale of cells is larger than the dissipation
scale.

Other relevant motivation comes from the non-gravitational heating
of baryon gas of clusters. In order to solve the deviation from the
similarity of virialized dark halos, various models of
non-gravitational heating and cooling of baryonic gas have been
proposed
\citep[e.g.][]{1999A&A...347....1V,2001ApJ...546...63T,2002ApJ...576..601V,
2003ApJ...588..704Z,2003ApJ...584...34X,2007ApJ...668....1N}. Since
these cooling and heating may introduce characteristic scales, the
self-similar hierarchy will no longer work on these characteristic
scales. Therefore, it would be worth to detect the scale on which
the above-mentioned two predictions to be broken.

We study these properties with the relation between the Compton
parameter $y$ of SZ effect and X-ray luminosity $L_{\rm x}$. The
thermal SZ effect is due to the inverse Compton scattering of cosmic
microwave background (CMB) photons by hot electrons of baryon fluid.
The Compton parameter $y$ depends on the pressure of electron gas
\citep{1969Ap&SS...4..301Z,1980ARA&A..18..537S}. There are
many works on the $y$-$L_x$ relation \citep[e.g.][]{2004MNRAS.348.1401D,
2007ApJ...668..772M,2007arXiv0708.0815B}. We will, however, focus on the
above-mentioned two points, which have not yet been addressed right
now.

The outline of this paper is as follows. \S 2 presents the scaling relations
$y=10^{A(r)}L_x^{\alpha(r)}$ with observed samples, and shows that $A(r)$ and
$\alpha(r)$ are scale-invariant. \S 3 describes the hydrodynamic cosmological
simulation samples. The comparison of the scaling relations of simulation
samples with observed results is presented in \S 4. The conclusions and
discussion are given in \S 5.

\section{$y$-$L_{\rm x}$ Scaling Relations from Observed Samples}

\subsection{Data}

To study the scale free properties, we should find the  $y(r)$ -
$L_{\rm x}(r)$ relations, where $y(r)$ and $L_{\rm x}(r)$ are,
respectively, the Compton parameter of SZ effect and X-ray
luminosity measured from regions with spatial scale $r$. The data of
X-ray luminosity of these clusters are taken from \cite{2003ApJ...591..526M}
(Xray1) and \cite{2007MNRAS.379..518M} (Xray2). The X-ray
luminosity from area on comoving scale $r$ is calculated by
\begin{equation}
L_{\rm x}(r)=\int_0^{\theta_{\rm r}}L_x(\theta)\theta {\rm d}\theta,
\end{equation}
where $\theta_{\rm r}=r/[(1+z)d_A(z)]$ is the angular radius
corresponding to the comoving scale $r$, and $d_A(z)$ is angular
diameter distance. $L_x(\theta)$ is proportional to the X-ray
surface brightness $S_x(\theta)$, which can be well fitted by
$\beta$-model $S_x(\theta)=S_{X0}[1+(\theta/\theta_{\rm
c})^2]^{(1-6\beta)/2}$ up to $\theta \sim 10$ arcmin.

Similarly, The mean of $y$ within a region on comoving scale $r$ is
given by
\begin{equation}
y(r)=\frac{2}{\theta_{\rm r}^2}\int_0^{\theta_{\rm r}}y(\theta)
\theta {\rm d}\theta.
\end{equation}
We will use the SZ effect data from \cite{2002ApJ...581...53R} (SZ1) and
\cite{2006ApJ...647...25B} (SZ2). The former compiled SZ effects of 18
clusters of galaxies spanning the redshift range of $0.14 < z <
0.78$, and the later includes 38 clusters in the same redshift
range. These data are on angular scales up to $\sim 2$ arcmin, of
which the corresponded $r$ is on about the same scale as, or larger
than, the Jeans length on redshift $\sim 0.5$. Moreover, the
$\theta$-dependencies of $y(\theta)$ are well fitted by
$\beta$-model $y(\theta)=y_0[1+(\theta/\theta_{\rm
c})^2]^{(1-3\beta)/2}$. Therefore, it would be reasonable to use
the $\beta$ model fitted $y(r)$ to study the $y(r)$ - $L_{\rm x}(r)$
relation. We will check this point below.

\subsection{Result}

Figure 1 plots the relation of $y(r)$ vs. $L_{\rm x}(r)$ on scales
$r$=0.1, 0.2, 0.39, 0.78, 1.56 and 3.12$h^{-1}$ Mpc respectively. In
this figure the SZ and X-ray data are taken from SZ1 and Xray1,
respectively. The cluster A370 is excluded as it shows a 3-$\sigma$
discrepancy with the distance-redshift relation \citep{2002ApJ...581...53R}.
Three clusters, Cl0016, A611 and A697, are also excluded due to lacking
the data of X-ray luminosity, and after all, there are totally 14
clusters used in Figure 1.

\begin{figure}
\includegraphics[width=14cm]{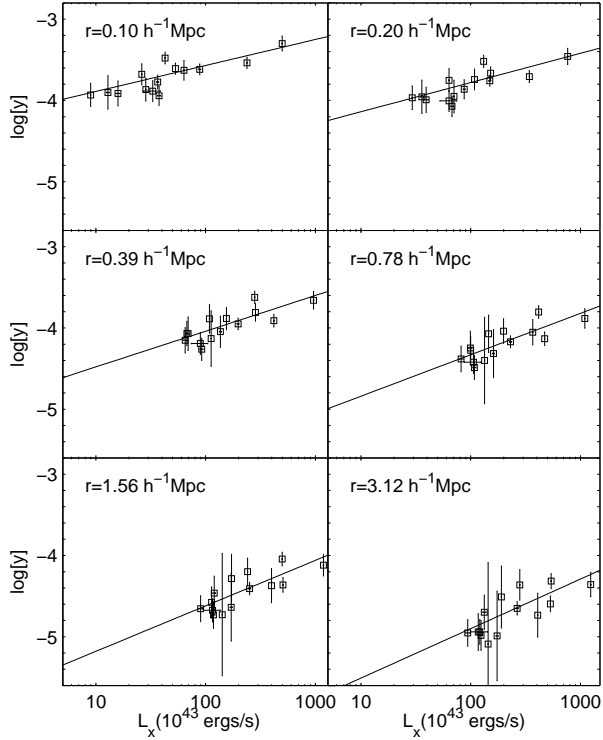}
\caption{$y$-$L_{\rm x}$ relation of observational samples SZ1+Xray1
(see Table 1). The Compton parameter $y$ are given by average over
areas on the comving
sizes 0.10, 0.20, 0.39, 0.78, 1.56, and 3.12 $h^{-1}$ Mpc,
respectively. The solid lines indicate the best-fitting for all
observational samples.}
\end{figure}

\begin{center}
\begin{tabular}{ccc}
\multicolumn{3}{c}{Table 1. $r$-dependence of $\alpha(r)$ and $A(r)$} \\[5pt]
\hline\hline
$r$($h^{-1})$ Mpc & $\alpha$ & $A$ \\
\hline
 0.10 & 0.32$\pm$0.06 & -4.21$\pm$0.11 \\
 0.20 & 0.36$\pm$0.08 & -4.50$\pm$0.17 \\
 0.39 & 0.44$\pm$0.10 & -4.92$\pm$0.23 \\
 0.78 & 0.51$\pm$0.11 & -5.35$\pm$0.27 \\
 1.56 & 0.56$\pm$0.12 & -5.74$\pm$0.30 \\
 3.12 & 0.61$\pm$0.13 & -6.12$\pm$0.33 \\
\hline
\end{tabular}
\end{center}

We make a best-fitting of the $y(r)$-$L_{\rm x}(r)$ relation with a
power law $y=10^{A(r)}L_{\rm x}^{\alpha(r)}$ for various scales $r$
as displayed in Figure 1. The coefficients $\alpha(r)$ and $A(r)$
are listed in Table 1. Both $\alpha(r)$ and $A(r)$ are significantly
dependent on the scale $r$. The exponent $\alpha(r)$ increases with
scale $r$, while amplitude $A(r)$ decreases with $r$. If the system
is given by a similar mapping of virialized halos, the scaling
relation should be $y\propto L_x^{3/4}$ \citep{1988MNRAS.233..637C},
which means that $\alpha$ is scale-independent and equal to 0.75.
Table 1 shows that the values of $\alpha$ on all scales are less
than 0.75. Accordingly, the baryon fluid on those scales should
dynamically deviate from a similar mapping of underlying virialized
gravitational field of dark matter halos. To check the effect of the
angular scales of $\sim 2$ arcmin (\S 2.1), we re-calculate the
$y(r)$-$L_{\rm x}(r)$ relation with clusters having $d_A>$1000 Mpc,
which with $r>0.5$ Mpc for angular scales 2 arcmin. We find that the
coefficients $\alpha(r)$ and $A(r)$ are consistent with Table 1 within
1-$\sigma$ range.

\begin{figure}
\centering
\includegraphics[width=14cm]{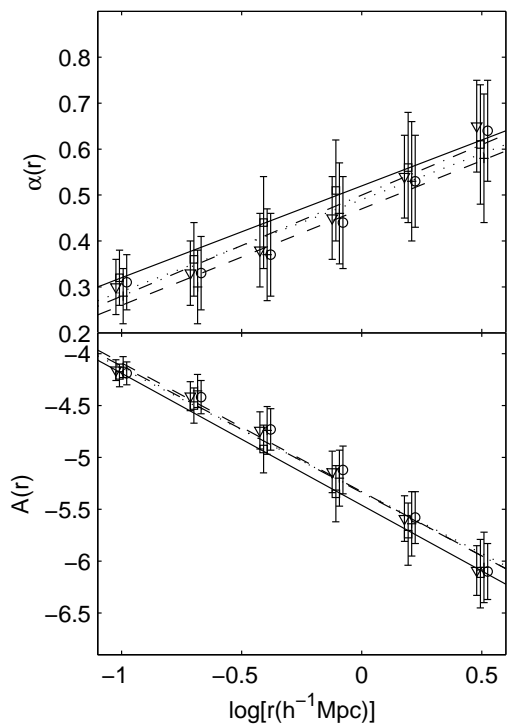}
\caption{The scale-dependence of the coefficients $\alpha(L)$ and
$A(r)$ for various observational samples:  SZ1+Xray1 (solid);
SZ1+Xray2 (dashed); SZ2+Xray1 (dotted); and SZ2+Xray2 (dot-dashed).}
\end{figure}

Figure 2 shows $\alpha(r)$ and $A(r)$ as functions of $r$. Note,
both $\alpha(r)$ and $A(r)$ can be well fitted, respectively by
$a_{\alpha}+b_{\alpha}\log r$ and $a_{A}+b_{A}\log r$, and
therefore, both $\alpha(r)$ and $A(r)$ are scale-invariant. The
amplitude $10^{A(r)}$ actually is a power law of $r$. This result
is consistent with the dynamics of self-similar hierarchy. To further
test this result, we used other data sets of SZ effect and X-ray
luminosity. The results are also shown in Figure 2. Although these
data sets are generally different from each other, all results of
$\alpha(r)$ and $A(r)$ can well be fitted by the straight line of
$\log r$. The fitting parameters are listed in Table 2. They are the
same within 1-$\sigma$ errors. It strongly supports the scenario of
scale-free dynamics.

\begin{center}
\begin{tabular}{ccccc}
\multicolumn{5}{c}{Table 2. Fitting Results of $\alpha(r)$ and $A(r)$.} \\[5pt]
\hline\hline
sample\footnotemark[1]
&  $a_{\alpha}$ & $b_{\alpha}$ &$a_{A}$ & $b_{A}$ \\
\hline
SZ1+Xray1 & 0.52$\pm$0.05 & 0.20$\pm$0.07 & -5.46$\pm$0.13 &-1.27$\pm$0.17 \\
SZ1+Xray2 & 0.47$\pm$0.06 & 0.21$\pm$0.08 & -5.33$\pm$0.13 &-1.24$\pm$0.17 \\
SZ2+Xray1 & 0.49$\pm$0.05 & 0.20$\pm$0.07 & -5.33$\pm$0.11 &-1.19$\pm$0.15 \\
SZ2+Xray2 & 0.50$\pm$0.04 & 0.22$\pm$0.06 & -5.34$\pm$0.10 &-1.22$\pm$0.13 \\
\hline
\multicolumn{5}{l}{\footnote[1]  SZ1 and SZ2 from \cite{2002ApJ...581...53R}
and \cite{2006ApJ...647...25B}, Xray1} \\
\multicolumn{5}{l}{ and Xrya2 from \cite{2003ApJ...591..526M}
(Xray1) and \cite{2007MNRAS.379..518M},}\\
\multicolumn{5}{l}{ respectively}.
\end{tabular}
\end{center}

It should also be pointed out that the scales of $r \leq 0.39$
$h^{-1}$ Mpc are actually less than the Jeans length of
baryon fluid. However, the coefficients $\alpha(r)$ and $A(r)$ are
still following the self-similar straight lines. In addition,
$\alpha(r)$ is seen to be less than the value 0.75 from virialized
halos. It implies that the dynamics of baryon fluid on these scales
seems still to be scale free.

\section{SZ effect samples of cosmological hydrodynamical simulations}

\subsection{Simulation}

Before embarking on the numerical calculations, we give a brief
summary on the baryon fluid when it is in the turbulence-like or
self-similar hierarchical clustering. The dynamics of growth modes
of clustering in an expanding universe is sketched by a stochastic
force driven Burgers equation \citep{1994PhRvL..72..458B,
1999MNRAS.307..376J,2002MNRAS.329...37M}. The turbulence-like behavior
is due to Burgers' turbulence, which will be developed when the Burgers
Reynolds number is large \citep{1995PhRvE..52.6183P,2004PhRvL..93r4503B,
1968PhFl...11..265K,2000PhRvL..84.2618L}. A turbulent flow in incompressible
fluid consists of vortexes, while the clustering of cosmic matter is
irrotational, because the modes with vorticity of the perturbed mass
density field do not grow. The Burgers' turbulence made the
initially random field to result in a collection of shocks and a
smooth variation of the field between the shocks. For cosmic baryon
fluid, the Burgers Reynolds number generally is larger in nonlinear
regime \citep{2004ApJ...612...14H}. Therefore, the baryon fluid in nonlinear
regime consists of a collection of shocks with various strengths in
both high, moderate and even low density areas. The kinetic energy of
fluid is dissipated due to the shocks on various scales. The kinetic
energy of fluid is effectively converted into thermal energy
\citep{2006PhRvL..96e1302H}. In this context, it is clear that the method of
hydrodynamical simulations should be capable of capturing
discontinuities, like shocks, and their effects on energy conversion
precisely.

We will use the Weightly Essentially NonOscillatory (WENO) algorithm, 
which is effective to capture shocks
and other discontinuities of the baryon fluid, and to give precise
value of the fluid field between the discontinuities
\citep{2004ApJ...612....1F}. This algorithm has been tested with 1.)
Shock tube; 2.) the Sedov-Taylor self-similar blast wave solution,
or the Bertschinger's similarity solution; 3.) Zeldovich pancake. It
is effectively to produce the baryon mass density contour, baryon
temperature contour around massive halos
\citep{2004ApJ...612....1F,2004ApJ...612...14H,2005ApJ...623..601H}.
Some other properties of this simulation algorithm can also be
found in \cite{2004ApJ...612....1F} and \cite{2004ApJ...612...14H}.

The simulations are performed in a cubic box 100$^3$ $h^{-3}$
Mpc$^3$ with a 1024$^3$ grid, and the number of dark matter
particles is 512$^3$. The mass of the dark matter particle is
$\sim 10^9$M$_{\odot}$, which corresponds to a density resolution
about $0.01$ times of the mean density of intergalactic medium (IGM). 
The grid size is then $100/1024\sim0.10$
$h^{-1}$ Mpc. This scale is smaller than the Jeans length at $z \leq
1$ \citep{2003ApJ...598....1B}. Therefore, the sample is suitable to
describe the baryon fluid from the dissipation scale to a few ten
$h^{-1}$ Mpc. We use the concordance $\Lambda$CDM cosmology model
with parameters $\Omega_{\rm m}$=0.27, $\Omega_{\rm b}$=0.044,
$\Omega_\Lambda$=0.73, $h$=0.71, $\sigma_8$=0.84, and spectral index
$n=1$ and the ratio of specific heats is $\gamma=5/3$. The transfer
function is calculated using CMBFAST \citep{1996ApJ...469..437S}.

The atomic processes including ionization, cooling and heating are
modeled as the method in \cite{1998MNRAS.301..478T}. We take a
primordial composition of H and He ($X$=0.76, $Y$=0.24) and use an
ionizing background model of \cite{2001cghr.confE..64H}.

The simulations start at the redshift $z=99$, and output the
density, velocity and temperature fields at redshifts $z$=2, 1, 0.5,
and 0. It is easy for hydrodynamic simulation with Eulerian variable
to reach low-density regions. They are suitable for a uniform
analysis of weakly as well as strongly clustered fields of baryon
fluid. These samples have been successfully applied to reveal the
self-similar hierarchical behavior of cosmic baryon fluid
\citep{2005ApJ...625..599K,2006PhRvL..96e1302H,2008ApJ...672...11L},
to explain the transmitted flux of HI and HeII Ly$\alpha$ absorption of
quasars \citep{2006ApJ...645..861L}, and to study the relations between X-ray
luminosity and temperature of groups of galaxies \citep{2006ApJ...642..625Z}.

As mentioned in \S 1, the self-similar hierarchical scenario
predicts that the relation $y(r)=10^{A(r)}[L_{\rm
x}(r)]^{\alpha(r)}$ given by areas containing collapsed objects
should also be available for areas without collapsed objects, only
if $r$ is larger than the dissipation scale. On the other hand,
either the effects of heating by injecting hot gas or metal cooling
are localized in massive halos. Their characteristic scales are less
than the Jeans length. Therefore, one can expect that the $y-L$
relations given by observed samples should be the same as that given
by simulation samples either with or without considering the
localized heating and cooling processes. To consider the effect of 
the metal abundance, we also use a sample with metal cooling of 
\cite{2006ApJ...642..625Z}. For this sample, The metal cooling
and metal line emission is calculated by the phenomenological
method: 1) assuming an uniform evolving metallicity $Z=0.3
Z_{\odot}(t/t_0)$, $t_0$ being the present universe age; 2)
computing the cooling function using the table of 
\cite{1993ApJS...88..253S}.

\subsection{Samples}

When relativistic corrections is negligible, the Compton parameter
$y$ of the thermal SZ effect along a line of sight, $l$, with an
angular distance $\theta$ from the center of a cluster in the plane
of the sky is given by
\begin{equation}
y(\theta)=\frac{k_{\rm B}\sigma_{\rm T}}{m_{\rm e}c^2}
    \int n_{\rm e}(l, \theta)T_{\rm e}(l,\theta){\rm d}l,
\end{equation}
where $\sigma_{\rm T}$ is the cross section of the Thomson
scattering; $n_{\rm e}$ and $T_{\rm e}$ are, respectively, the
number density and temperature of hot electrons. Since H and He
atoms are almost fully ionized, we take the electron density $n_{\rm
e}=\rho/\mu_{\rm e}m_{\rm p}$ where $\rho$ is the density of baryon
gas, $\mu_{\rm e}=2/(1+X)$ with a hydrogen abundance of $X=0.76$.

Using simulated density and temperature fields, we calculate the
parameter $y(\theta)$ with Eq.(3). The mean of parameter $y(r)$ on
various scales can be obtained using the scaling function of the
discrete wavelet transform (DWT)
\begin{equation}
y_{\bf j,l} = \frac{1 } {\int\phi_{\bf j,l}({\bf x}){\rm d}{\bf x} }
  \int y({\bf x})\phi_{\bf j,l}({\bf x}){\rm d}{\bf x}.
\end{equation}
where $\phi_{\bf j,l}({\bf x})$ is the scaling functions related to
cell $({\bf j,l})$ with comoving size $100/2^j$ $h^{-1}$ Mpc and at
position ${\bf l}=(l_1,l_2,l_3)$. The details of the DWT analysis
can be found in \cite{Fang&Thews1998}. We take the comoving size
$100/2^{j}$ $h^{-1}$ Mpc with $j=$10, 9, 8, 7, 6 and 5,
corresponding to scales 0.10, 0.20, 0.39, 0.78, 1.56 and 3.12
h$^{-1}$ Mpc, which are the same as that used in the analysis of the
observed samples in last section.

The total X-ray luminosity of thermal bremsstrahlung emission from a cell
$({\bf j,l})$ is calculated by the same way as \cite{2006ApJ...642..625Z}
\begin{equation}
(L_{\rm x})_{\bf j,l}= V_{j}\frac{1 } {\int\phi_{j,l}({\bf x}){\rm
d}{\bf x} }
  \int \epsilon^{ff}({\bf x})\phi_{j,l}({\bf x}){\rm d}{\bf x},
\end{equation}
where $\epsilon^{ff}({\bf x})$ is the map of X-ray emissivity, and
$V_j$ is the volume of cell $({\bf j,l})$.

The DWT decomposition has following advantages. First, the set of
scaling functions are orthogonal and complete, and therefore, the
decomposition of Eqs.(3) and (4) does not give rise to false
correlation. Second, the DWT cell $({\bf j,l})$ with high mass
density can directly be used to identify clumpy structures. The
DWT-identified cells on scale 1.5 $h^{-1}$ Mpc is statistically the
same as clusters identified by traditional method, such as the
friend-of-friend algorithm \citep{1998ApJ...508..472X}. Therefore, the DWT
variables $y_{\bf j,l}$ and $(L_{\rm x})_{\bf j,l}$ provides a
uniform description of the Compton parameter and X-ray luminosity of
the whole field, regardless of the dynamical details of all cells
$({\bf j,l})$. Third, the DWT variables can be applied directly to
non-Gaussian behavior \citep{2006PhRvL..96e1302H,2008ApJ...672...11L}
We will use the Harr wavelet (Daubechies 2) to do the calculation below.
We also repeat the calculations with wavelet Daubechies 4. The
statistical features given by Daubechies 4 are basically the same
as Haar wavelet.

\section{Scaling Relations between the SZ Effect and X-ray Luminosity}

\subsection{$y$-$\rho_{\rm igm}$ and $L_x$-$\rho_{\rm igm}$ relations}

Figure 3 shows the relations between the mean Compton parameter
$y(r)$ and mean mass density $\rho_{\rm igm}(r)$ of cells on scale
$r$ for the simulation samples at redshift $z=0.5$. The comoving
scales $r$ are taken to be 0.39, 0.78, 1.56 and 3.12 $h^{-1}$ Mpc
respectively. Figure 3 has a dark area as a bottom envelop of the
$y$-$\rho_{\rm igm}$ distribution. It gives a tight correlation
between $y$ and $\rho_{\rm igm}$ and can be described approximately
by a power law of $y \propto \rho_{\rm igm}^{1.8}$ for all scales
$r$. This relation is basically consistent with the so-called
adiabatic `equation of state' $T \propto \rho^{2/3}$ if considering
$y\propto \rho_{\rm igm} T$.

\begin{figure}
\includegraphics[width=14cm]{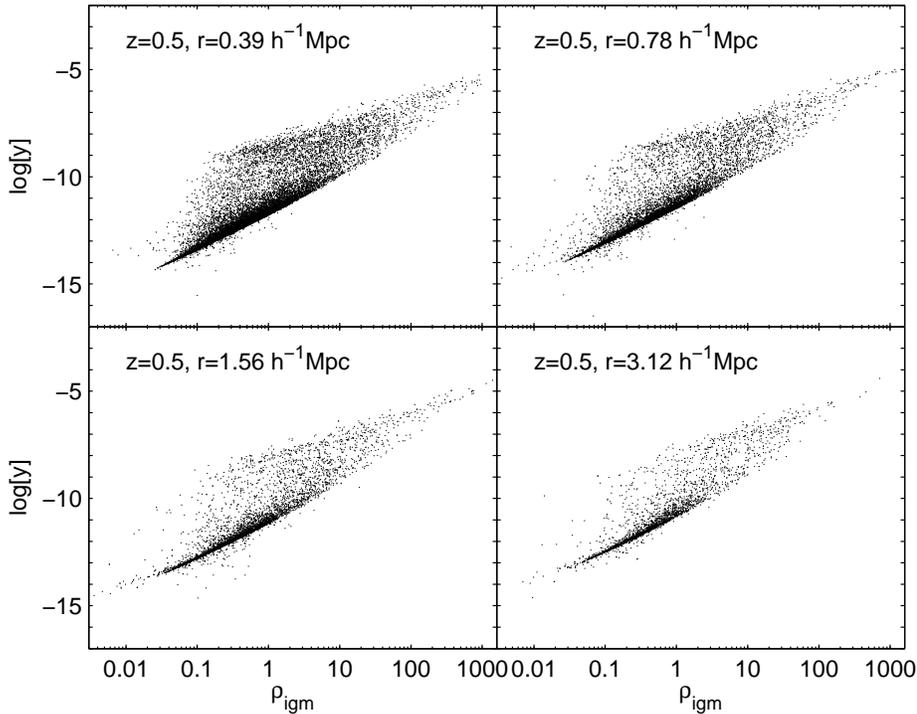}
\caption{Compton parameter $y$ vs. baryon density $\rho_{\rm igm}$
of simulation samples at redshift $z=0.5$, in which $y$ and
$\rho_{\rm igm}$ are the mean over cells with comoving scales
$r=0.39$, 0.78, 1.56, and 3.12 $h^{-1}$ Mpc, respectively.
$\rho_{\rm igm}$ is in unit of mean mass density $\bar{\rho}_{\rm
igm}$ of the field.}
\end{figure}

In Figure 4, the similar analysis has been performed for the
relations between the total X-ray luminosity $L_x(r)$ and mean mass
density $\rho_{\rm igm}(r)$. Clearly, there is also a dark area as
the bottom envelop of the $L_X$-$\rho_{\rm igm}$ distribution. The
tight correlation of the dark area yields $L_X\propto \rho_{\rm
igm}^{2.4}$ for all scales. This is expected if considering $L_X
\propto \rho^2_{\rm igm} T^{1/2}$.

\begin{figure}
\includegraphics[width=14cm]{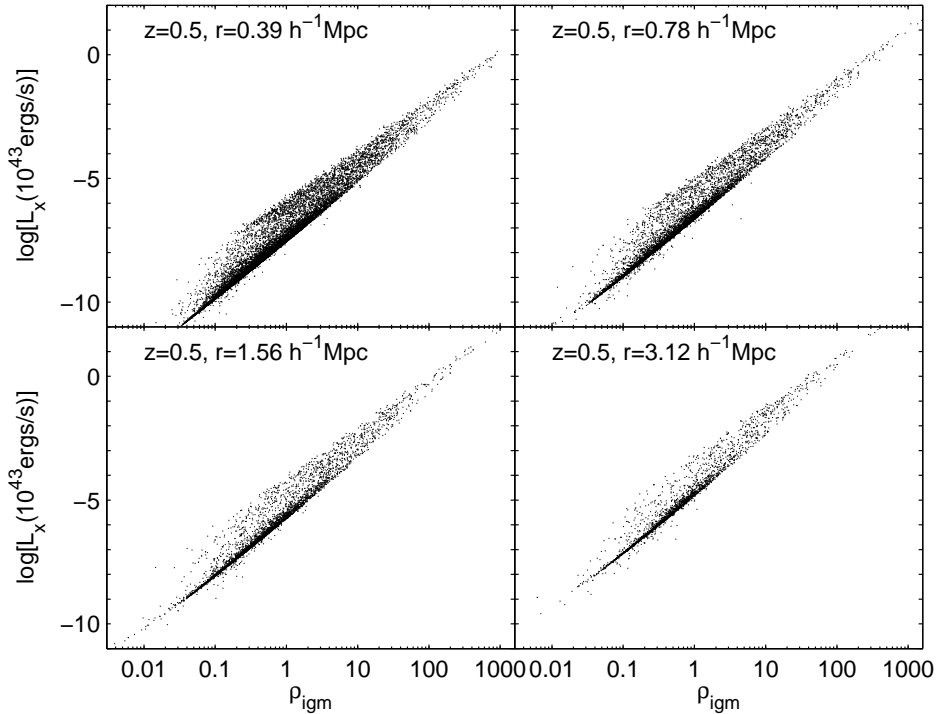}
\caption{X-ray luminosity $L_x$ vs. baryon density $\rho_{\rm igm}$
of simulation samples at redshift $z=0.5$. The $L_x$ is the total
X-ray luminosity from cell with comoving scale $r=0.39$, 0.78, 1.56,
and 3.12 $h^{-1}$ Mpc. $\rho_{\rm igm}$ are the mean of density
within the cell. $\rho_{\rm igm}$ is in unit of mean mass density
$\bar{\rho}_{\rm igm}$ of the whole field.}
\end{figure}

Therefore, $y$ and $L_X$ given by the tight correlations of bottom
envelops in Figures 3 and 4 imply $y\propto L^{0.75}$, which is the
same as that given by a virialized underlying gravitational field.
If we adopt the adiabatic `equation of state' $T\propto \rho^{2/3}$,
the $y$-$L_{\rm x}$ scaling relation should be $y\propto
L_x^{0.71}$. The observed coefficient $\alpha$ are less than 0.75 on
all scales. The reason for $\alpha < 0.75$ is clearly shown by
Figures 3 and 4. For a given $\rho_{\rm igm}$, the distribution of
$y$ is significantly scattered from the bottom envelop. For sample
at $z\simeq 0.5$, the data points of $y$ corresponding to $\rho_{\rm
igm} \simeq 1$ scatter in the range from $\sim 10^{-11}$ to
$10^{-8}$. The scattering is due to the heating of Burgers' shocks,
which leads to the baryon fluid to be multiphasic, i.e. the relation
between temperature and mass density can not be described by {\it
one} polytropic equation. The points of $y$ higher than the bottom
envelop correspond to state with temperature higher than those
given by the `equation of state' $T \propto \rho^{2/3}$
\citep{2004ApJ...612...14H,2005ApJ...625..599K}.

The multiphases of baryon fluid are seen for all scales. The
scattering in the $y$-$\rho_{\rm igm}$ distribution is substantial
in high mass density areas ($\rho_{\rm igm}>1$) as well as low mass
density areas ($\rho_{\rm igm}<1$). The $L_x$-$\rho_{\rm igm}$
distribution (Figure 4) shows similar scattering. It can be
understood that the Burgers' shock heating is not only working in
high density regions, but also in low density areas. It leads to the
deviation of $\alpha$ from 0.75.

\subsection{$y$-$L_{\rm x}$ scaling relation}

We now study the scaling relation between $y(r)$ and $L_{\rm x}(r)$.
Figure 5 plots the distribution of the Compton parameter $y(r)$
against the X-ray luminosity $L_{\rm x}(r)$ for the simulation
samples at $z\leq 0.5$, in which the comoving scales of the cell are
taken to be 0.78, 1.56 and 3.12 $h^{-1}$ Mpc respectively. The
observed points of $y(r)$-$L_{\rm x}(r)$ are taken from the samples
SZ1+Xray1.

\begin{figure}
\includegraphics[width=14cm]{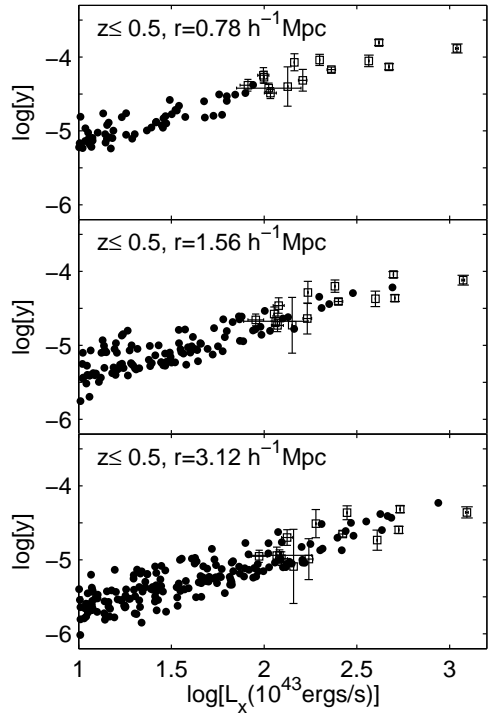}
\caption {The scaling relation between the Compton parameter $y(r)$
and X-ray luminosity $L_{\rm X}(r)$. The scales $r$ are 0.78, 1.56,
and 3.12 $h^{-1}$ Mpc. The observed data points (squares with
errorbars) are from SZ1+Xray1 and the simulation data points
(dotted) are taken from sample of redshift $z\leq 0.5$.}
\end{figure}

On the scale of 3.12 $h^{-1}$ Mpc, the simulation data can be fitted
by observations very well. Using the simulation data with
$\log[L_X(r)/10^{43}{\rm erg/s]}\geq 1$, we found that the best
fitting scaling relation $y=10^{A}L_x^{\alpha}$ of simulation sample
gives $\alpha=0.65\pm0.03$ and $A=-6.31\pm 0.05$, which are the same
as observed samples shown in Table 1. For the scale of 1.56 $h^{-1}$
Mpc, Figure 5 also shows the consistency between observed and
simulated samples. Since $\alpha$ is less than 0.75, the scaling
relation shown in Figure 5 is dominated by data points located above
the bottom envelop of Figure 3. In other word, the $y$-$L_{\rm x}$
relations are mainly determined by the structures which are involved
in the evolution of the Burgers turbulence.

On scale 0.78 $h^{-1}$ Mpc, the simulation sample still shows the
same trend as observed data, but there are fewer points with $\log[
L_X(r)/10^{43}{\rm erg/s}]\geq 2$. This phenomenon is more serious
on scale 0.39 $h^{-1}$ Mpc. In this case, the best fitting scaling
relation yields $\alpha=0.65\pm0.01$ and $A=-5.47\pm0.01$, which
deviates from the observed result $\alpha=0.44\pm0.10$ and
$A=-4.92\pm0.23$ (Table 1). The deviation on scales
$\leq 0.78$h$^{-1}$Mpc is expected. Since the scale 0.78 $h^{-1}$
Mpc is typical of the so-called $R_{2500}$ of clusters, and it is
less than the virialization radius of clusters. Within this scale
range, dissipation and non-gravitational processes are involved. It is
beyond the regime of scale-free evolution.

Because the observed data contains clusters with redshifts higher
than 0.5, we also made a comparison between observed data and
simulation samples of $z\leq 1$. The results are displayed in Figure
6, which yields almost the same results as Figure 5. We also analyzed
the simulation samples including metal cooling of \cite{2006ApJ...642..625Z}.
The result is given in Figure 7. It shows the effect of metal cooling
does not change the feature of Figure 5. Although metal cooling may
have a big effect on $L_{\rm x}$ for groups. But it will have the
similar effect on $y$. The $y$-$L{\rm x}$ scaling relation still keeps
self-similar in the inertia range.

\begin{figure}
\includegraphics[width=14cm]{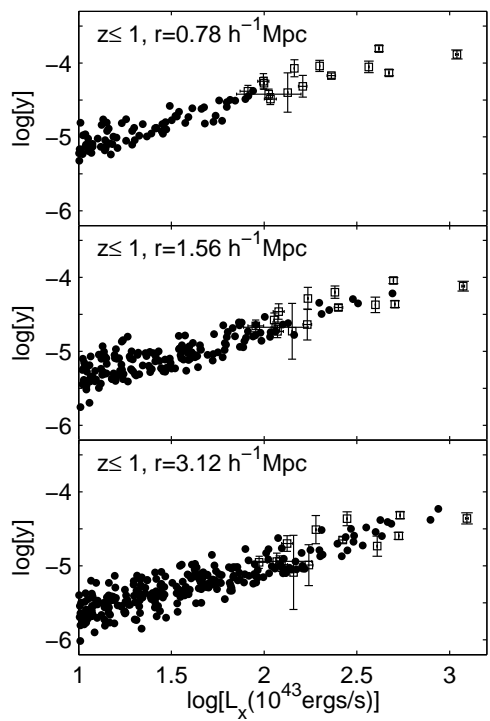}
\caption {The scaling relation between the Compton parameter $y(r)$
and X-ray luminosity $L_{\rm X}(r)$. The scale $r$ are 0.78, 1.56,
and 3.12 $h^{-1}$ Mpc respectively. The observed data points (
squares with errorbars) are from SZ1+Xray1. The simulation data
points (dots) are taken from sample of redshift $z\leq 1$.}
\end{figure}

\begin{figure}
\includegraphics[width=14cm]{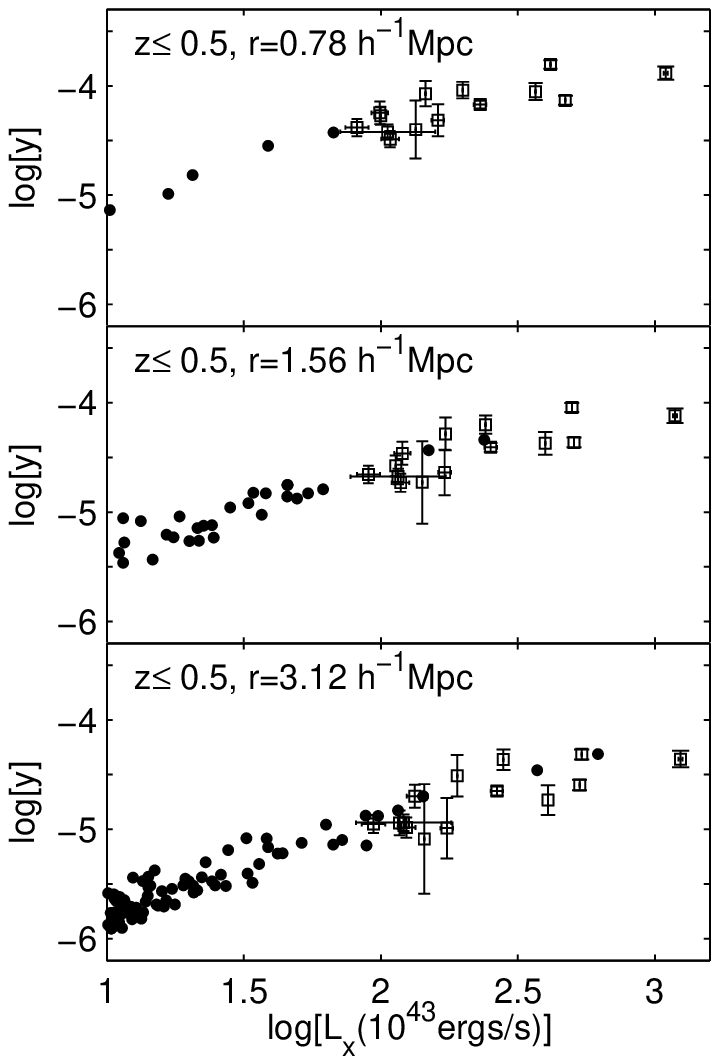}
\caption {The same as Figure 5, but with simulation sample in which the
the metal cooling and metal line emission are considered.}
\end{figure}

\subsection{Available range of the $y$-$L_{\rm x}$ scaling relation}

As emphasized above, the scaling relation between physical
quantities of cosmic baryon fluid due to the Burgers turbulence
should hold for the entire field. Those relations inferred from
statistical analysis made in regions containing collapsed
structures, like clusters, should also be applicable in regions
without that structures. Actually, Figures 5 - 7  have already
shown that the $y$-$L_x$ scaling relation works well for all regions
containing strong X-ray emission $\log[L_X(r)/10^{43}{\rm
erg/s}]\geq 1$.

Figure 8 presents the $y$-$L_x$ scaling relation covering a much
wider range of the X-ray luminosity $-3 \leq \log[L_X(r)/10^{43}{\rm
erg/s}]\leq 3$, from rich clusters to weakly clustered areas. It
shows clearly that the scaling relation of Figure 1 is available in
the weakly clustered areas.

\begin{figure}
\includegraphics[width=14cm]{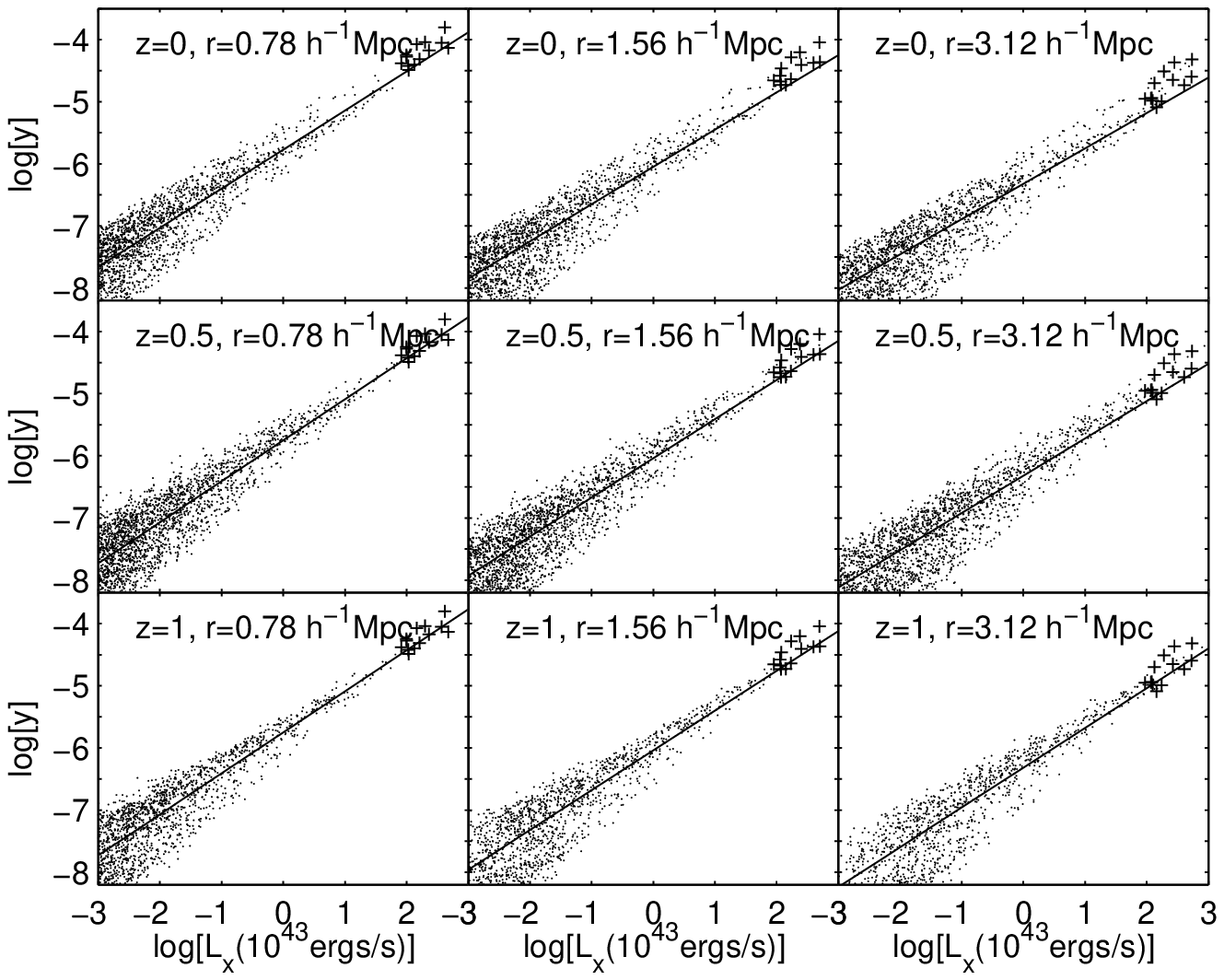}
\caption {The scaling relation between the Compton parameter $y$ and
X-ray luminosity $L_{\rm x}$ of simulation data at redshift
$z=0$ (top), 0.5 (middle), and 1 (bottom) respectively. The scale $r$
are 0.78 (left), 1.56 (middle) and 3.12 (right) $h^{-1}$ Mpc
respectively. The solid line is given by the fitting of simulation
data.}
\end{figure}

\begin{center}
\begin{tabular}{ccc}
\multicolumn{3}{c}{Table 3. Fitting Coefficients
$\alpha$ and $A$ for Simulation Data}\\[5pt]
\hline\hline
 range of$\log[L_X(r)]$  & $\alpha$  & $A$ \\
\hline
  1 $<\log[L_X(r)/10^{43}{\rm erg/s}]<$ 3 & 0.65$\pm$0.03 & -6.31$\pm$0.05 \\
  0 $<\log[L_X(r)/10^{43}{\rm erg/s}]<$ 3 & 0.63$\pm$0.01 & -6.27$\pm$0.02 \\
 -1 $<\log[L_X(r)/10^{43}{\rm erg/s}]<$ 3 & 0.65$\pm$0.01 & -6.29$\pm$0.01 \\
 -2 $<\log[L_X(r)/10^{43}{\rm erg/s}]<$ 3 & 0.65$\pm$0.01 & -6.30$\pm$0.01 \\
 -3 $<\log[L_X(r)/10^{43}{\rm erg/s}]<$ 3 & 0.59$\pm$0.00 & -6.32$\pm$0.01 \\
\hline
\end{tabular}
\end{center}

We did a fitting of $y$-$L_x$ scaling relation in the simulation
sample on the scale of $r=3.12$ $h^{-1}$ Mpc and $z\leq0.5$ with
different ranges of $\log L_X(r)$. The best-fitting values of the
coefficients $\alpha$ and $A$ for various ranges of $\log L_X(r)$
are listed in Table 3. Obviously, the coefficients $\alpha$ and $A$
are almost independent of the range of $\log L_X(r)$. The scaling
relation, $y(r)=10^{A(r)}[L_X(r)]^{\alpha(r)}$ is very stable within
$10^{39}<L_x<10^{46}$ erg s$^{-1}$. If $y$ and $L_x$ are measured
from regions on scales larger than dissipation scales, the $y$-$L_x$
scaling relation still holds regardless of the dynamical details on
the dissipation scales. The wide range of $L_{\rm x}$ is consistent
with the scenario of self-similar hierarchical evolution, which
gives a unified description of the dynamics of clustering on various
level in the scale-free range.

Similar analysis was made in simulation samples on scale of
$r=1.56h^{-1}$ Mpc and 0.78 $h^{-1}$ Mpc at redshifts $z=0$, 0.5 and
1 respectively. These scaling relations are also valid within the
entire range of $10^{39}<L_x<10^{46}$ erg s$^{-1}$ as well.
Moreover, from Figure 8, we can see that the scaling relations
basically are redshift-independent within the range $z\leq 1$. Along
with the decreasing redshift, there are more data points with larger
$\log L_X$.  The $y$-$L_x$ correlation shown in Figure 8 can be seen
as a tree with root at left-bottom corner, and tip at right-top
corner. The formation of clustered objects leads to the growth of
the tip of the tree along the direction given by the scaling
relation $y(r)=10^{A(r)}[L_x(r)]^{\alpha(r)}$.

\section{Discussion and Conclusions}

The scaling studied in this paper comes from the scale-free dynamics of 
cosmic baryon fluid in the expanding universe. That is, there is no 
preferred special scales can be identified in the range from the onset 
of nonlinear evolution down to the characteristic length of dissipation.
The clustering evolution is described by self-similar hierarchy of baryon 
fluid. This self-similar hierarchy is different from the gravitational 
self-similarity of virialzed system, which is characterized by the mass 
of the system. On the other hand, the scaling from the scale-free dynamics 
is characterized by the scale range, on which physical quantities of the 
mass and velocity field of cosmic baryon fluid are measured. That is, 
the dimensional quantities measured in cells with size larger than the 
dissipation characteristic length should satisfy the same scaling 
relations, regardless whether the cells contains massive collapsed 
objects. The scaling will be broken if the scale is less than the 
characteristic length of dissipation.  

We demonstrated this scaling with the $y(r)$-$L_{\rm x}(r)$ relation of 
baryon field. The observed $y(r)$-$L_{\rm x}(r)$ relation can be well 
reproduced with the hydrodynamical simulation. The important point is 
that the fitting of $y(r)$-$L_{\rm x}(r)$ scaling is not based on the 
identified clusters in simulation samples, but using all cells on a 
given scale $r$, regardless whether the cell contains rich clusters 
with strong X-ray emission. In other words, the scaling relations 
$y(r) =10^{A(r)}[L_{\rm x}(r)]^{\alpha(r)}$ can be used to
describe cells containing strong X-ray emission, $L_{\rm x}>10^{43}$ erg
s$^{-1}$, as well as very weak X-ray emission regions $L_{\rm
x}\sim10^{40}$ erg s$^{-1}$. This result supports the self-similar
hierarchical scenario of the clustering of baryon fluid.

The observed $y(r)$-$L_{\rm x}(r)$ relations starts to show a deviation 
from simulation results on scales smaller than 0.78 h$^{-1}$ Mpc, which 
would be the characteristic scale of the dissipation. We calculated the 
Jeans lengths for each objects used in our statistics. All of the Jeans 
lengths are found to be less than 0.8 h$^{-1}$ Mpc. It is very well 
consistent with our result. Moreover, this result is also consistent
with previous studies on the non-gravitational heating of clusters. The 
non-gravitational heating generally is considered to be due to the 
injection of hot gas and energy from SN and AGN. However, the significant 
effect of injecting hot gas and energy by SN is mostly arising from 
dwarf galaxies, i.e. on scales much smaller than clusters. The Ly$\alpha$ 
observations of protoclusters shows that the feedback of AGN is not 
strong enough to heat the gas of clusters within comoving size 0.5 h$^{-1}$ 
Mpc \citep{2003ApJ...584...45A}. Therefore, the effect of the SN and AGN 
heating would be dramatic only on scales smaller than about 0.5 h$^{-1}$ 
Mpc, but probably cannot heat gas within cells on scales $\geq 1$ h$^{-1}$ 
Mpc to amount of the order of 1 keV per nucleon 
\citep{1999ApJ...510L...1P,1999ApJ...524...22W}. Actually we find that 
scaling relations on scales $>0.78$h$^{-1}$ Mpc are still held after 
subtracting the contributions from the central part ($<0.2$ h$^{-1}$ Mpc)
of clusters to both $L_{\rm x}$ and $y$, as done in 
\cite{1998ApJ...504...27M,2007ApJ...668..772M}.

As an application, the scaling relations of $y(r)$-$L_{\rm x}(r)$ would be
useful for estimating the contamination of SZ effect on CMB, which
is important, especially, on small scales \citep{2006MNRAS.369..645C}.
Recent simulation has shown that the Planck project would be capable of
probing $y$ on the order of $y=10^{-7}-10^{-8}$ \citep{2005MNRAS.363...29D}.
It might give a direct test on the universal scaling relations of
$y(r)$-$L_{\rm x}(r)$ given by the self-similar hierarchical evolution.

Our simulation of the baryon fluid is within the Eulerian framework.
One can also study the nature of intermittency of fluid with
Lagrangian point of view. In this approach, the hierarchical
clustering can be tracked with Lagrangian trajectories. It has been
found that the intermittent scaling is related to the long time
correlations in the particle acceleration, namely, the random forces
driving the particle motion is long range correlated
\citep{2002PhRvL..89y4502M}. It would be interesting to investigate the
scaling in the Lagrangian scheme.

\noindent{\bf Acknowledgments.}

T.-J.Z. is supported by the Fellowship of the World Laboratory.
H.-Y.W., T.-J.Z. and L.-L.F. acknowledge support from the National
Science Foundation of China (grants 10473002, 10573036 and
10545002). T.-J.Z thanks Xiang-Ping Wu for his valuable discussion.
This work was also partially supported by US NSF AST 05-07340.

\end{document}